\documentclass[conference]{IEEEtran}
\usepackage{cite}
\usepackage{amsmath,amssymb,amsfonts}
\usepackage{algorithmic}
\usepackage{graphicx}
\usepackage{textcomp}
\usepackage{xcolor}
\def\BibTeX{{\rm B\kern-.05em{\sc i\kern-.025em b}\kern-.08em
    T\kern-.1667em\lower.7ex\hbox{E}\kern-.125emX}}

\usepackage{bm, comment}
\usepackage{mathabx,fixmath, algorithmic}
\usepackage{pstool}
\usepackage{psfrag}
\usepackage{hyperref}

\def\x{\mathbold{x}}
\def\z{\mathbold{z}}
\def\Z{\mathbold{Z}}
\def\f{\mathbold{f}}
\def\y{\mathbold{y}}
\def\g{\mathbold{g}}

\def\q{\mathbold{q}}
\def\k{\mathbold{k}}
\def\K{\mathbold{K}}
\def\I{\mathbold{I}}
\def\Boeta{\mathbold{\eta}}
\def\bxi{\mathbold{\xi}}
\def\blambda{\mathbold{\lambda}}
\def\A{\mathbold{A}}
\def\B{\mathbold{B}}

\def\P{\mathbold{P}}

\def\Exp{\mathsf{E}}

\def\reals{\mathbb{R}}
\newcommand{\sym}[1]{\mathbb{S}^{#1}}
\def\transp{\intercal}
\def\conv{\mathsf{conv}}
\DeclareMathOperator*{\minimize}{\mathsf{minimize}}
\DeclareMathOperator*{\argmin}{\mathsf{argmin}}
\DeclareMathOperator*{\maximize}{\mathsf{maximize}}

\newtheorem{theorem}{Theorem}

\newtheorem{definition}{Definition}

\newenvironment{proof}{\emph{Proof:}}{\hfill $\blacksquare$}

\begin{document}

\title{Smooth Strongly Convex Regression}

\author{\IEEEauthorblockN{Andrea Simonetto}
\IEEEauthorblockA{\textit{IBM Research Ireland} \\
Dublin, Ireland \\
andrea.simonetto@ibm.com} \vspace*{-.75cm}
}

\maketitle

\begin{abstract}
Convex regression (CR) is the problem of fitting a convex function to a finite number of noisy observations of an underlying convex function. CR is important in many domains and one of its workhorses is the non-parametric least square estimator (LSE). Currently, LSE delivers only non-smooth non-strongly convex function estimates. In this paper, leveraging recent results in convex interpolation, we generalize LSE to smooth strongly convex regression problems. The resulting algorithm relies on a convex quadratically constrained quadratic program. We also propose a parallel implementation, which leverages ADMM, that lessens the overall computational complexity to a tight $O(n^2)$ for $n$ observations.  Numerical results support our findings. 
\end{abstract}


\section{Introduction}

Convex regression (CR) is concerned with fitting a convex function to a finite number of observations.
In particular, suppose that we are given $n$ noisy observations of a convex function $\varphi: \reals^d \to \reals$ as  
\begin{equation}
y_i = \varphi(\x_i) + \epsilon_i, \quad i \in I_n := \{1, \ldots, n\},
\end{equation}
where $\epsilon_i$'s are random variables, while $\x_i \in \reals^d$. The objective of CR is then to estimate the true function $\varphi$, given the observations $y_i$'s, in a way in which the estimated function $\hat{\varphi}$ is convex. 

CR is a particular class of shape-constrained regression problems, and since its first conception in the 50's, it has attracted much attention in various domains, such as statistics, economics, operations research, signal processing and control~\cite{Samworth2018, Groeneboom2001, Aybat2014}. In economics, CR has been motivated by the need for approximating consumers' utility functions from empirical data~\cite{Meyer1968}, a task which has been recently re-considered in the context of personalized optimization with user's feedback~\cite{Simonetto2019}. 

In this paper, we study least squares estimators (LSEs) for CR. LSEs have some key advantages over many other estimators proposed in the literature for CR (e.g., constrained Gaussian processes~\cite{Wang2016}, splines~\cite{Dontchev2003}, or others~\cite{Birke2007}). First, LSEs are non-parametric and hence they do not require any tuning and avoid the issue of selecting an appropriate estimation structure. Second, LSEs can be computed by solving a convex quadratic program. Therefore, at least in theory, they can be solved very efficiently using interior point methods. Third, being based on the least squares paradigm, they can be naturally extended to time-varying cases (when the function to be estimated changes continuously in time) by, e.g., exponential forgetting coefficients; these cases are becoming more and more important in the current data streaming era~\cite{SPM}.  

One of the main theoretic drawback is however that the class of functions that can be enforced is limited to the general convex functions, while in many applications one would like to be able to impose at least smoothness and/or strong convexity. 

In this paper, our contributions are as follows,

$\bullet$ First, we propose a novel smooth strongly convex CR algorithm. The resulting estimator has at its heart a convex quadratically constrained quadratic program, with $n+nd$ variables and $n (n-1)$ constraints. We also report on its computational complexity as a function of $n$. The building blocks for this novel algorithm are the recent results in smooth strongly convex interpolation~\cite{Taylor2016, Taylor2017}.

$\bullet$ Second, we propose a decomposition scheme based on the alternating direction method of multipliers (ADMM)  to lessen the computational complexity and make the method parallel. The resulting computational complexity per iteration is then $O(n^2)$ which is tight (i.e., no LSE can obtain a lower computational complexity). The ADMM approach is based on a properly constructed constraint graph, as well as a dual formulation of the local sub-problems.

The results presented in this paper generalize LSE to smooth strongly convex functions, and the non-smooth results can be re-obtained as a special case\footnote{
{\bf Notation.} Vectors are indicated with $\x \in \reals^n$, and matrices with $\A \in \reals^{m \times n}$. The Euclidean norm is indicated with $\|\x\|_2$, the infinity norm as $\|\x\|_{\infty}$. $(\cdot)^\transp$ is the transpose operator. Symmetric positive (semi)-definite matrices of dimension $n$ are indicated as $\A \in \sym{n}_{+} (\A \in \sym{n}_{++})$. For convex functions $\varphi: \reals^n \to \reals$, we indicate with $\partial \varphi (\x)$ their subgradient at point $\x$ and with $\varphi^\star$ their convex conjugate. 
$O(\cdot)$ is the standard big-O notation.
}.

\section{Definitions and problem statement}

We start by formally defining the functional class of interest. Given two parameters $\mu$ and $L$ satisfying $0 \leq \mu < L \leq +\infty$, we consider convex functions satisfying both a smoothness and a strong convexity condition. Given a  convex function $\varphi: \reals^d \to \reals$, we say that the function is $L$-smooth and $\mu$-strongly convex, which we denote with $\mathcal{F}_{\mu, L}$, iff the following two conditions are satisfied:
\begin{itemize}
\item Inequality $1/L \|\g_1 - \g_2\|_2 \leq \|\x_1 - \x_2\|_2$ holds $ \forall \x_1, \x_2 \in \reals^d$ and corresponding subgradients $\g_1, \g_2 \in \reals^{d}$;
\item Function $\varphi(\x) - \mu/2 \|\x\|^2_2$ is convex.
\end{itemize}
These definitions allow for $L$ to be equal to $+\infty$ (i.e., the non-smooth case). In the case of a finite $L$, the first condition implies differentiability of the function. When $L = +\infty$, this condition becomes vacuous, and the function can be non-differentiable. The class of generic convex functions simply corresponds to $\mathcal{F}_{0, +\infty}$. The case $L = \mu$ can be discarded, as it only involves simple quadratic functions, that can be estimated parametrically much more efficiently. 

The problem we are interested in solving can be then formalized by using the empirical $\ell_2$ norm as:
\begin{equation}\label{eq.inf}
\hat{\varphi}_n \in \argmin_{\psi \in \mathcal{F}_{\mu, L}}\Big\{ \sum_{i\in I_n} (y_i - \psi(\x_i))^2 \Big\} \,.
\end{equation}
As we will see, the solution of~\eqref{eq.inf} will consist of two parts: \emph{(i)} the solution of a finite-dimensional optimization problem defined on the observation set, and \emph{(ii)} an interpolating function which maintains the functional class also in all the other points of the domain. In this respect, we define the notion of $\mathcal{F}_{\mu, L}$-interpolation as follows. 

\smallskip 
\begin{definition}
The set $\{(\x_i, \g_i, f_i)\}_{i \in I_n}$ where $\x_i, \g_i \in \reals^d$ and $f_i \in \reals$ for all $i \in I_n$ is $\mathcal{F}_{\mu, L}$-interpolable iff there exists a function $\varphi \in \mathcal{F}_{\mu, L}$, $\varphi: \reals^d \to \reals$, such that $\g_i \in \partial \varphi(\x_i)$ and $f_i = \varphi(\x_i)$ for all $i \in I_n$. \hfill $\diamond$
\end{definition}

\section{Shape-constrained least-squares}

\subsection{State of the art}

The infinite dimensional optimization problem in~\eqref{eq.inf} can be reduced to a finite dimensional one for the case $\mathcal{F}_{0, +\infty}$ as follows. Let $f_i = \varphi_n(\x_i)$, for $i\in I_n$ and define the vector $\f = [f_1, \ldots, f_n]^\transp \in \reals^n$. Let $\g_i = \partial \varphi_n(\x_i)$, and define the vector $\g = [\g_1^\transp, \ldots, \g_n^\transp]^\transp \in \reals^{nd}$. 
We can now rewrite the optimization problem~\eqref{eq.inf} for $\mathcal{F}_{0, +\infty}$ on the observation set as the following convex quadratic program (QP):
\begin{subequations}\label{qp}
\begin{eqnarray}
\!\!\!\!\!\!\!\minimize_{\f \in \reals^n\!,~  \g\in\reals^{nd}} && \!\!\!\!\!\!\!\sum_{i\in I_n} (y_i - f_i)^2 \\
\!\!\!\!\!\!\!\mathsf{subject~to:} && \!\!\!\!\!\!\!f_j + \g_j^\transp (\x_i - \x_j)\leq f_i, \,\, \forall i, j \in I_n. \label{base-constr}
\end{eqnarray} 
\end{subequations}
Problem~\eqref{qp} is a convex QP with $n + nd$ variables and $n(n-1)$ constraints (after removing the trivial $i=j$ ones). Constraint~\eqref{base-constr} imposes convexity on the observation points.    
 
A solution of~\eqref{qp}, that can be labelled as $\{(f_i^*, \g_i^*)\}_{i \in I_n}$, 
is $\mathcal{F}_{0, +\infty}$-interpolable by construction. In fact, once the solution is retrieved, an allowed estimator/interpolating function for function $\varphi$ at point $\x \in \reals^d$ is given by
\begin{equation}\label{eq.ns}
\hat{\varphi}_n(\x) = \max_{i \in I_n} \left\{f_i^* +  \g_i^{*,\transp}(\x - \x_i) \right\}.
\end{equation}

Two comments are in order at this point: 

$\bullet$ First, estimator~\eqref{eq.ns} is non-smooth and non-strongly convex in general. While \emph{ad-hoc} smoothing techniques do exist~\cite{Mazumder2019}, one incurs a daunting trade-off between smoothing quality (in terms of low Lipschitz constant $L$) and estimation quality (in terms of small error $\varepsilon$ w.r.t. the non-smooth solution). Typically, if one wants to retrieve functions of the class $\mathcal{F}_{\mu, L}$ with a low $L$, one has to expect poor estimation quality $\varepsilon$. 

$\bullet$ Second, even though problem~\eqref{qp} is a convex QP and one can use off-the-shelf solvers to solve it efficiently (e.g., OSQP~\cite{osqp}, or ECOS~\cite{ecos}), the number of shape constraints is still $O(n^2)$. In addition, its computational complexity grows at least as $O(n^3(d+1)^3)$, making practically hard to solve problems with $n>200$ and even small $d$. However, thanks to the decomposable structure of problem~\eqref{qp} one can resort to first-order methods~\cite{Aybat2014, Mazumder2019, Lin2020} whose computational complexity scales as $O(n^2)$ per iteration, to tackle problems up to $n \sim O(1000)$ in dimensions that can go up to $d=200$ for the very recent~\cite{Lin2020}. Note that a computational complexity of $O(n^2)$ is the least one can expect, given the $O(n^2)$ constraints. Finally, partitioning methods have also been advocated~\cite{Hannah2013} but not explored here.

\subsection{Smooth strongly convex regression}

In this paper, we propose a new set of constraints (instead of~\eqref{base-constr}) together with an interpolation procedure for $\hat{\varphi}_n(\x)$ to enforce smoothness and strong convexity automatically. We use and adapt results from~\cite{Taylor2016, Taylor2017} for this purpose. The basic idea is that smoothness and strong convexity interchange via the procedure of convex conjugation. While we leave the technical details to the above mentioned papers, we can cite the following interpolability results. 

\smallskip
\begin{theorem}{\emph{ ($\mathcal{F}_{\mu,L}$-interpolability) \cite[Theorem~4]{Taylor2016}}}
The set $\{(\x_i,\g_i, f_i)\}_{i\in I_n}$ is $\mathcal{F}_{\mu,L}$-interpolable iff the following set of conditions holds for every pair of indices $i, j \in I_n$:
\begin{multline}\label{int.constr}
f_i - f_j - \g_j^\transp (\x_i - \x_j) \geq \\\frac{1}{2(1 - \mu/L)}\left(\frac{1}{L}\|\g_i - \g_j \|^2_2 + \mu \|\x_i - \x_j\|_2^2 \right. \\ \left. - 2 \frac{\mu}{L} (\g_j-\g_i)^\transp (\x_j - \x_i) \right).
\end{multline}
\end{theorem}

\smallskip




With this in place, we are ready to modify the constraint set \eqref{base-constr} to accommodate $\varphi \in \mathcal{F}_{\mu, L}$. In particular, to solve~\eqref{eq.inf} for a $\varphi \in \mathcal{F}_{\mu, L}$, we can leverage the convex problem:  
\begin{subequations}\label{socp}
\begin{eqnarray}
\minimize_{\f\in\reals^n\!,~ \g\in\reals^{nd}} && \sum_{i\in I_n} (y_i - f_i)^2 \\
\mathsf{subject~to:} && \eqref{int.constr}, \quad \forall i, j \in I_n.
\end{eqnarray} 
\end{subequations}
This is a convex quadratically constrained quadratic problem (QCQP), a special case of a second-order conic program~\cite{Boyd2004a}. Once a solution $\{(f^*_i, \g_i^*)\}_{i \in I_n}$ is found, the following theorem describe an allowed estimation/interpolation strategy.  

\smallskip
\begin{theorem}\label{th.interpolation}
For any set $\{(\x_i,\g_i^*, f_i^*)\}_{i\in I_n}$ that is $\mathcal{F}_{\mu,L}$-interpolable, an allowed interpolating function is
\begin{equation}\label{interp}
\hat{\varphi}_n(\x) = \conv(p_i(\x)) + \frac{\mu}{2} \|\x\|^2_2
\end{equation}
where \vspace*{-.2cm}
\begin{multline}
p_i(\x) := \frac{L-\mu}{2} \| {\x} - \x_i\|_2^2 + (\g_i^*-\mu \x_i)^\transp \x + \\ - \g_i^{*,\transp} \x_i + f_i^* + \mu/2\|\x_i\|_2^2,
\end{multline}
and where $\conv(\cdot)$ indicates the convex hull.
\end{theorem}

\smallskip
\begin{proof}
See Appendix~A.
\end{proof}

\smallskip
Problem~\eqref{socp} together with the interpolation strategy~\eqref{interp} yield the promised smooth strongly convex estimator for function $\varphi$. If we set $\mu = 0$ and $L = +\infty$, we retrieve the non-smooth estimator. For $\mu = 0$, $L< +\infty$, we obtain a non-strongly convex smooth estimator, while for $\mu > 0$ and $L = +\infty$ a non-smooth strongly convex one.

Problem~\eqref{socp} is a QCQP in $n$+$nd$ variables and $n(n$-$1)$ constraints, which can be solved with off-the-shelf convex solvers, e.g., ECOS~\cite{ecos}, or MOSEK. Since the computational complexity grows at least as $O(n^3 (d+1)^3)$, similar practical limitations than the non-smooth estimator apply here. We will show how to overcome them by resorting to ADMM next.


\subsection{Parallel implementation}

Since the computational complexity of~\eqref{socp} could be prohibitive for practical applications, we move now to understand how one can decompose the problem into smaller parts and reduce the overall complexity. Strategies to use first-order methods (proximal methods and ADMM) on the quadratic problem~\eqref{qp} for non-smooth convex functions have been reported in~\cite{Aybat2014,Mazumder2019,Lin2020}. In the case of~\eqref{qp}, the problem is separable in both cost and constraints (once dualized) and a decomposition strategy is rather direct. For the case of~\eqref{socp} however, each constraint couples the variables $\g_i$ and $\g_j$ due to the quadratic term, and a  decomposition is not immediate. We consider here a novel edge-based ADMM decomposition. 

Consider the set of constraints of type~\eqref{int.constr} for all $i,j \in I_n$ and define the constraint graph, as a \emph{directed} graph $G = (V, E)$, whose vertices are the nodes $i \in I_n$, and edges are \emph{all the  combinations} of $i,j$: $E = \{(i,j)| i\in I_n,\, j \in I_n,\, i \neq j\}$. The cardinality of $E$ is $|E|= n (n-1)$. For each edge $e\in E$, consider its two nodes, say $i$ and $j$, and define the edge variables $\Boeta_{e,i} = [f_i^e, \g_i^{e,\transp}]^\transp$, $\Boeta_{e,j} = [f_j^e, \g_j^{e,\transp}]^\transp$, as well as  $\bxi_{e} = [\Boeta_{e,i}^\transp, \Boeta_{e,j}^\transp]^\transp \in \reals^{2(1 + d)}$. In this context, each directed edge has its own functional and derivative variables. 

We use the notation $i \sim e$ to indicate that node $i$ is one of the two vertices of edge $e$, while we explicitly write $e(i\to j)$ to indicate that edge $e$ is the directed edge with $i$ as source node and $j$ as sink node.  For ease of representation, we also define the local constraint, 
\begin{multline}\label{int.constr.admm}
\mathcal{C}_{e(i \to j)} := \big\{ f_i^e - f_j^e - \g_j^{e, \transp}(\x_i - \x_j) \geq \\\frac{1}{2(1 - \mu/L)}\left(\frac{1}{L}\|\g_i^e - \g_j^e \|^2_2 + \mu \|\x_i - \x_j\|_2^2 \right.\\\left. - 2 \frac{\mu}{L} (\g_j^e-\g_i^e)^{\transp} (\x_j - \x_i) \right)\Big\}.
\end{multline}

We now split problem~\eqref{socp} at the edges, so that each variable $\Boeta_{e,i}^\transp$ containing node $i$, is different for each edge $e$ having node $i$ has one of its vertices. Then we enforce equality of all the node variables via the supporting vector $\z_i \in \reals^{1+d}$ for each $i$, $\z = [\z_1^\transp, \ldots, \z_n^\transp]^\transp \in \reals^{n+nd}$. With this philosophy, problem~\eqref{socp} can be rewritten in the equivalent form:
\begin{subequations}\label{socp-admm}
\begin{eqnarray}
\!\!\!\!\!\!\minimize_{\bxi \in \reals^{2|E|(1+d)}\!,~ \z\in\reals^{n+nd}} && \frac{1}{2n}\sum_{e \in E} \, \sum_{i\sim e} \, (y_i - f_i^e)^2 \\
\!\!\!\!\!\!\mathsf{subject~to:} && \mathcal{C}_{e(i \to j)}, \,\,\,\, \forall e \in E,  \\
&& \Boeta_{e,i} = \z_{i}, \,\,\,\, \forall e \in E,\, i \sim e. \label{constr.a}
\end{eqnarray} 
\end{subequations}
It is key that the constraint $\mathcal{C}_{e(i \to j)}$ is present only once for each directed edge.  

The above problem can now be tackled with ADMM. We leave the derivations out, since standard, and report only the final result. 
Start with some initialization for the auxiliary variable $\z$ and initialize the \emph{scaled} dual variables of the constraints~\eqref{constr.a} as $\blambda = [\ldots, \blambda_{e,i}^\transp, \blambda_{e,j}^\transp, \ldots]^\transp \in \reals^{2 |E| (1 + d)} = {\bf 0}$. Set the penalization $\rho>0$. Then at each step,

\begin{enumerate}
\item Solve the edge QCQP for each edge $e \in E$:
\begin{multline}\label{loc.qcqp}
\!\!\!\!\!\!\!\![\Boeta_{e,i}^{+,\transp}, \Boeta_{e,j}^{+,\transp}]^\transp = \bxi^{+}_e = \argmin_{\bxi_e \in \mathcal{C}_{e(i \to j)}} \Big\{\frac{1}{2n}\sum_{i \sim e} (y_i - f_i^e)^2 + \\  + \sum_{i \sim e} \frac{\rho}{2} \left\|\Boeta_{e,i}  - \z_i + \blambda_{e,i}\right\|_2^2,
\Big\}
\end{multline} 
\item Update the $\z_i$ variables for each node $i \in V$:
\begin{equation}
\z_i^{+} = \frac{1}{2 n} \sum_{e| i\sim e} \Boeta_{e,i}^+ 
\end{equation}
\item Update the $\blambda_e$ variables $\forall e \in E$, and node $i \sim e$:
\begin{equation}
\blambda_{e,i}^+ = \blambda_{e,i} + (\Boeta_{e,i}^+ - \z_i^{+})
\end{equation}
\item Set $\bxi_e = \bxi^{+}_e$ for all $e \in E$, $\z_i = \z_i^+$ for all $i \in V$, and $\blambda_{e,i} = \blambda_{e,i}^+$, for all $e \in E$, $i \sim e$, and go to 1).
\end{enumerate}

Convergence of the above procedure to a minimizer of the original problem~\eqref{socp} is assured by the standard ADMM convergence results~\cite{Boyd2011}. In practice, one stops the ADMM iterations after a specified error criterion has been met, or a number of iterations has been reached. In this case, we consider as approximate solution to our problem, the final $\z$ vector that ADMM yields (we let $\tilde{f}_i$ be the near-optimal $\f$ values and $\tilde{\g}_i$ be the near-optimal $\g$ values) and we use the approximate interpolating function [see Th.~\ref{th.interpolation}]:
\begin{equation}\label{interp-1}
\hat{\varphi}_n(\x) = \conv(\tilde{p}_i(\x)) + \frac{\mu}{2} \|\x\|^2_2,
\end{equation}
\vspace*{-.5cm}
\begin{multline}
\tilde{p}_i(\x) := \frac{L-\mu}{2} \| {\x} - \x_i\|_2^2 + (\tilde{\g}_i-\mu \x_i)^\transp \x + \\ - \tilde{\g}_i^\transp \x_i + \tilde{f}_i + \mu/2\|\x_i\|_2^2.
\end{multline} 
Note that function~\eqref{interp-1} is still a $L$-smooth $\mu$-strongly convex function by construction, just the ADMM approximate solution $\z$ is only approximately $\mathcal{F}_{\mu, L}$-interpolable, which means that $\tilde{f}_i \approx \hat{\varphi}_n(\x_i)$ and $\tilde{\g}_i \approx \partial\hat{\varphi}_n(\x_i)$ instead of equality (meaning that function~\eqref{interp-1} is only approximately minimizing the LS metric). In practice, we notice that this approximation delivers good feasible results for small errors. 


\subsection{Solving the local dual}

Solving~\eqref{loc.qcqp} in the primal form is still too computational complex (especially if it has to be done for each edge for each iteration of the ADMM algorithm). However, \emph{since there is only one scalar constraint}, the dual is mono-dimensional and easier to solve with a few iterations of a projected Newton's method. To do this, we write~\eqref{loc.qcqp} in the standard form:
\begin{subequations}
\label{loc.qcqp.st}
\begin{eqnarray}
\minimize_{\bxi_e\in\reals^{2(1+d)}} &&  \bxi_e^\transp \P_0 \bxi_e + \q_0^\transp \bxi_e + r_0 \\
\mathsf{subject~to:} && \bxi_e^\transp \P_1 \bxi_e + \q_1^\transp \bxi_e + r_1 \leq 0,
\end{eqnarray} 
\end{subequations}
where $\P_0 \in \sym{2(d+1)}_{+}$, $\P_1 \in \sym{2(d+1)}_{++}$, $\q_0 \in \reals^{2(d+1)}$, $\q_1 \in \reals^{2(d+1)}$, $r_0 \in \reals$, and $r_1 \in \reals$ are properly defined matrices, vectors, and scalars to match \eqref{loc.qcqp.st} to the original \eqref{loc.qcqp}. Define the Lagrangian function
\begin{equation}
\mathcal{L}(\nu) := \bxi_e^\transp \P_0 \bxi_e + \q_0^\transp \bxi_e + r_0 + \nu (\bxi_e^\transp \P_1 \bxi_e + \q_1^\transp \bxi_e + r_1),
\end{equation}
for the dual variable $\nu\geq 0$. Note that $\P_0$ is positive definite by construction (due to $\rho>0$), and therefore setting $\P_{\nu} :=\P_0 + \nu \P_1$, the inverse $\P_{\nu}^{-1}$ exists and it is well-defined. Set also $\q_{\nu} := \q_0 + \nu \q_1$. The dual problem of~\eqref{loc.qcqp.st} is:
\begin{equation}\label{dual}
\maximize_{\nu \geq 0} \phi(\nu) := -\frac{1}{4}\q_\nu^\transp \, \P_\nu^{-1}\, \q_\nu + \nu r_1 + r_0
\end{equation}
whose solution can be found with standard projected Newton's method\footnote{
For the sake of completeness, the gradient of the dual function is
\begin{equation*}
\frac{\textrm{d} \phi}{\textrm{d}\nu} = -\frac{1}{2} \q_1^\transp \P_{\nu}^{-1}\q_{\nu} + \frac{1}{4} \q_{\nu}^\transp \P_{\nu}^{-1}\P_1  \P_{\nu}^{-1}\q_{\nu} + r_1,
\end{equation*}
while the Hessian is
\begin{equation*}
\frac{\textrm{d}^2 \phi}{\textrm{d}\nu^2} = -\frac{1}{2} \q_1^\transp \P_{\nu}^{-1}\q_{1} +  \q_{1}^\transp \P_{\nu}^{-1}\P_1  \P_{\nu}^{-1}\q_{\nu} -\frac{1}{2} \q_{\nu}^\transp  \P_{\nu}^{-1}\P_1  \P_{\nu}^{-1}\P_1  \P_{\nu}^{-1} \q_{\nu}. 
\end{equation*}
}.
Once the dual problem is solved and the unique dual solution $\nu^*$ is found, the unique primal solution of~\eqref{loc.qcqp.st} can be retrieved as
\begin{equation}
\bxi^*_{e} = -\frac{1}{2} \P_{\nu^*}^{-1}\, \q_{\nu^*}. 
\end{equation}

Putting things together, the ADMM procedure (1-4) with a a Newton's method to solve the mono-dimensional local dual problems~\eqref{dual} has computational complexity of $n (n-1) \times O((d+1)^3)$ (where the $O((d+1)^3)$ term comes from the computation of the inverse $\P_{\nu}^{-1}$) and for $d \ll n$, the leading error is $O(n^2)$.

\section{Numerical evaluation}

We evaluate the presented algorithms in terms of computational time and mean square error on a toy problem. The aim is to evaluate scalability and quality in a simple one-dimensional setting. A more detailed numerical investigation is deferred to future research efforts. All the computations are performed using Python 3.6, on a 2.7 GHz Dual-Core Intel Core i5 laptop with 8GB of RAM. We use CVXPY~\cite{cvxpy} for solving the non-smooth and smooth problems. Internally, the QPs are solved with OSQP, while the QCQPs with ECOS. 

The true generating function is $\varphi(x) = x^2$ over $x \in [-1,1]$. Function $\varphi$ is $2$-smooth and $2$-strongly convex. The observations are generated by adding noise drawn from $\mathcal{N}(0, \sigma^2)$.   

We run \emph{(i)} the non-smooth problem with QP~\eqref{qp} and estimator~\eqref{eq.ns}; \emph{(ii)} the $L$-smooth $\mu$-strongly convex problem with QCQP~\eqref{socp} and estimator~\eqref{interp}; \emph{(iii)} the edge-based ADMM to solve the smooth problem with local problems~\eqref{loc.qcqp} and approximate estimator~\eqref{interp-1}, with $\rho = 1/n$ and initial $\z$ specified running a Gaussian process estimator~\cite{Rasmussen2006}, which offers a computationally cheap smooth (but in general non-convex) first approximation
. For the Gaussian process we use a square exponential kernel. For the ADMM we solve the local problems resorting to their duals and employing a projected Newton's method with back-tracking. The stopping criterion for the ADMM has been selected as
\begin{equation}
\varepsilon = \max\{\|[\Boeta_{e,i}^+ - \z_i^+]_{e\in E, i\sim e}\|_{\infty}, \|\z^{+} - \z \|_{\infty}\}.
\end{equation}

\begin{figure}
\centering
\includegraphics[width = .5\textwidth, trim = {0, .5cm, 0cm, .5cm}, clip]{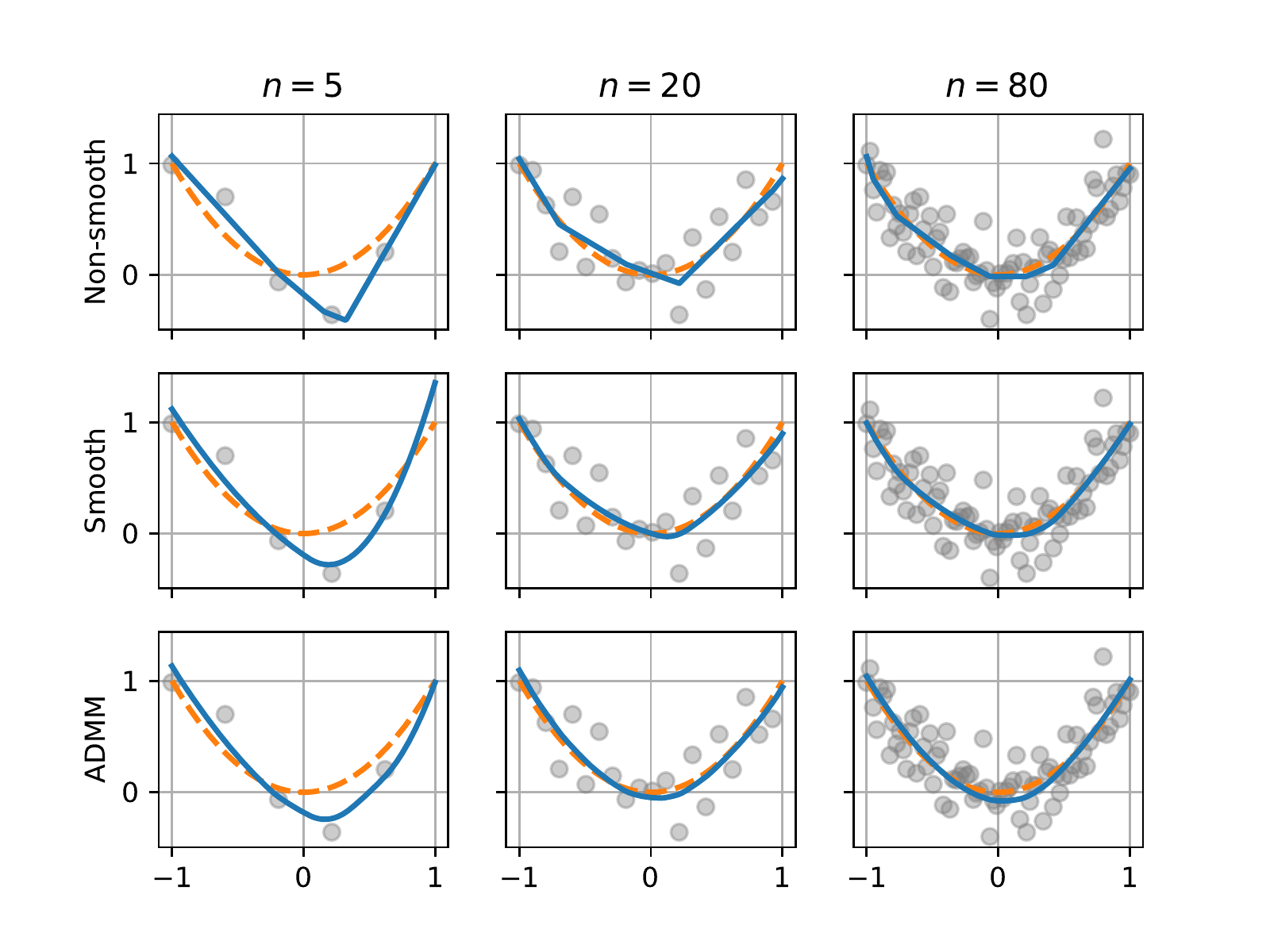}
\caption{Examples of estimation of the convex function $\varphi(x) = x^2$ with $n$ noisy observations. In continuous blue line the estimated $\hat{\varphi}$, in dashed orange line the true function. The observations are indicated with grey dots.}
\label{fig.1}
\end{figure}

In Figure~\ref{fig.1}, we report the three methods (non-smooth, smooth, and ADMM) for the task of estimating the given function $\varphi(x)$ with an increasing amount of observations. Here the observation noise standard deviation is set to $\sigma = 0.2$, while $\varepsilon = 0.01$. We also set $\mu = 1$ and $L=5$. We display the estimated function $\hat{\varphi}(x)$ with a continuous blue line, while the true function is a dashed orange line. Using grey dots, we represents the observations. As we notice, non-smooth and smooth estimators achieve the desired estimation task with higher accuracy the more observations are present. ADMM also does well (even with a moderate error $\varepsilon$). 

We run simulations for different $n$'s, considering $\sigma = 0.1$, and two versions of ADMM, one with $\varepsilon_1 = 0.03$ and the other with $\varepsilon_2 = 0.01$. (All the other parameters have been left the same as the ones in Figure~\ref{fig.1}). All the results are averaged over $10$ realizations of the observations for the centralized problems, and $25$ for the parallel methods (which are more dependent on the realization for the number of iterations).

\begin{figure}
\centering
\vspace*{-.25cm}
\includegraphics[width = .5\textwidth, trim = {0, .0cm, 0cm, .5cm}, clip]{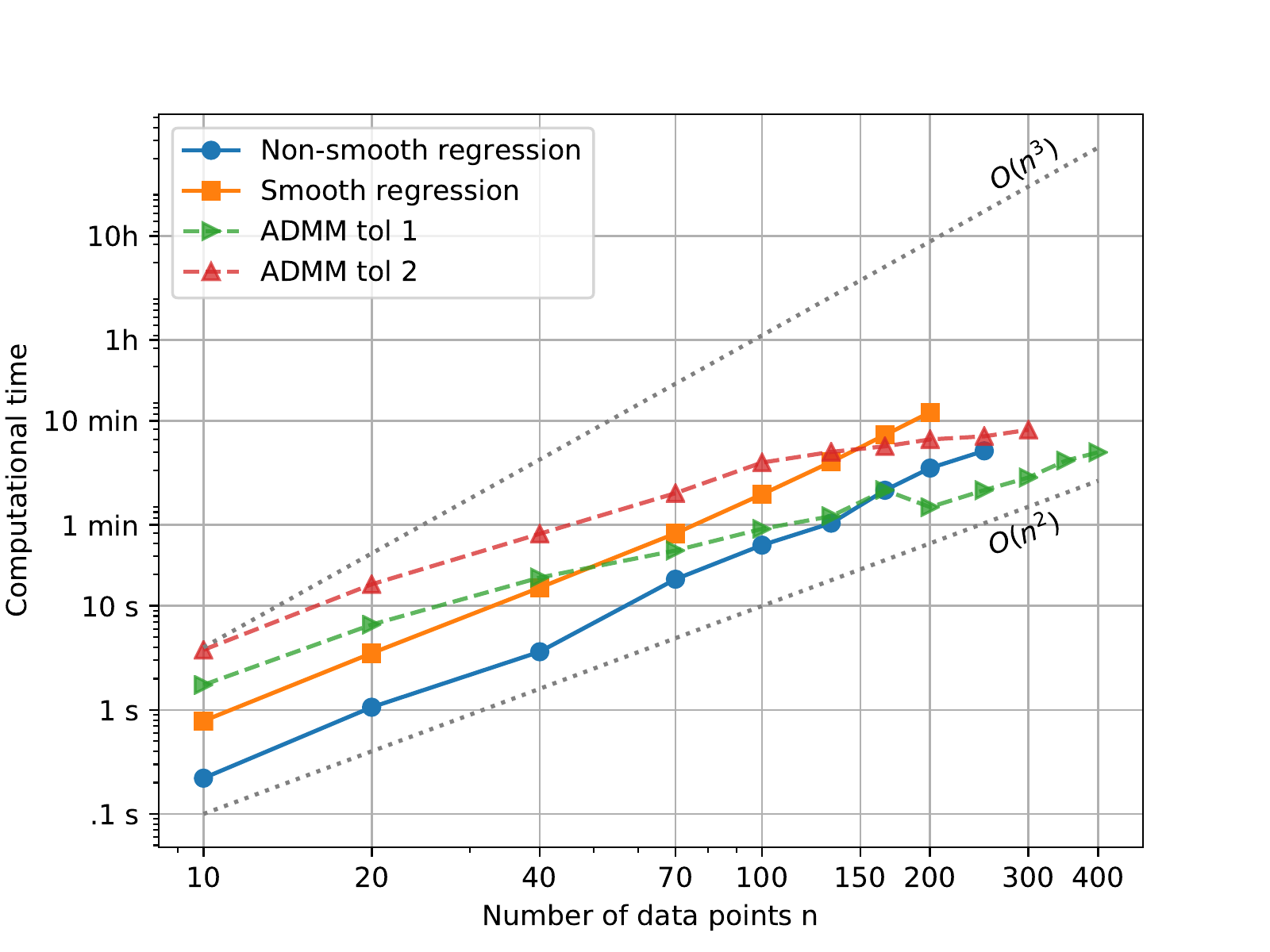}
\caption{Computational time for the different methods varying $n$. The dotted lines indicate the $O(n^2)$ and $O(n^3)$ growth.}
\label{fig.2}
\end{figure}

Figure~\ref{fig.2} represents the computational times of the various methods. As we see, the smooth method is the slowest one (as $n$ increases), as expected. The ADMM methods run slower than the other methods at the beginning due to the non-accurate initialization (and therefore due to the need for more iterations to reach the specified stopping criterion), while as $n$ increases, they run the fastest (since the initialization becomes better and better).

\begin{figure}
\centering
\vspace*{-.25cm}
\includegraphics[width = .5\textwidth, trim = {0, .0cm, 0cm, .5cm}, clip]{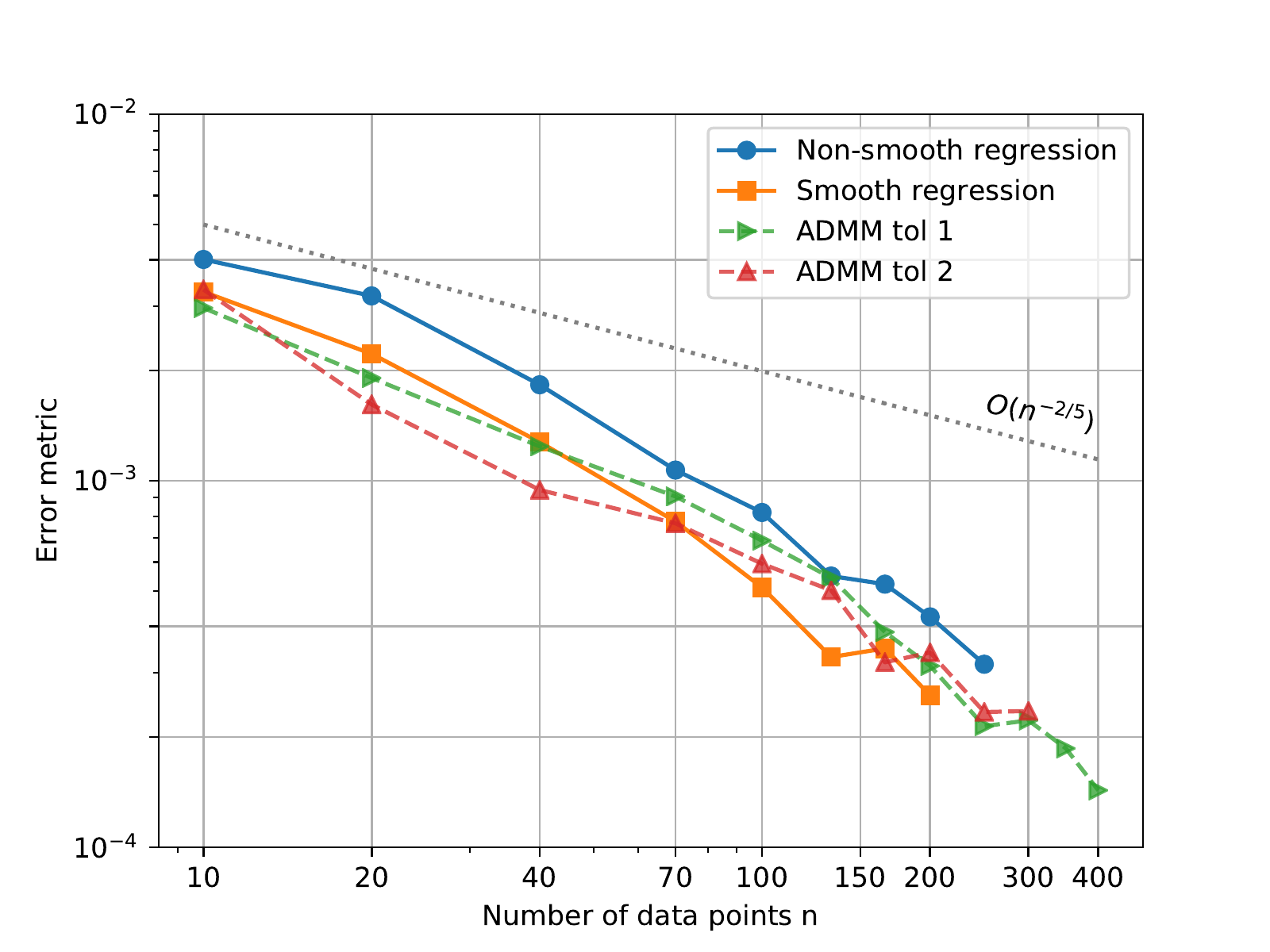}
\caption{Mean square error metric for the estimator $\hat{\varphi}_n$ for the different methods varying $n$ (cf.~\eqref{mse}). The dotted line indicates the theoretical $O(n^{-2/5})$ convergence rate for the mean square error metric on the observation points. }
\label{fig.3}
\end{figure}

Figure~\ref{fig.3} represents an error metric defined as 
\begin{equation}\label{mse}
E_{\hat{\varphi}_n} = \frac{1}{n_s}\sum_{s\in I_s} (\hat{\varphi}_n(y_s) - \varphi(y_s))^2
\end{equation}
This metric is defined over a finer equi-spaced sampling $y_s, s = \{1, \dots, n_s\} =: I_s$, over $[-1,1]$ with $n_s = 1000$. The idea is to capture the error in the estimator $\hat{\varphi}_n$, rather than just the mean square error on the observation points. As we see and expected, the smooth estimator is better than the non-smooth one. ADMM delivers estimates of comparable accuracy than the smooth one. In the figure, we also display an $O(n^{-2/5})$ line, which is the typical convergence rate of non-parametric LSE for $d=1$~\cite{Seijo2011, Lim2012, Blanchet2019, Mazumder2019}, in the mean-square-error-on-the-observation-points sense. 

What is interesting here is that the metric $E_{\hat{\varphi}_n}$ is built on the estimated $\hat{\varphi}_n$ which for ADMM is the approximate~\eqref{interp-1}. Because of this, the ADMM's estimator can be better than the smooth one (especially in low observation settings) This is per se quite interesting and will be further explored in the future.




\section{Conclusion}

We have proposed a method to perform smooth strongly convex regression in a non-parametric least squares sense. The method relies on the solution of a convex quadratically constrained quadratic program. We have discussed computational complexity and offered a first-order alternative based on ADMM to lessen the computational load to a tight $O(n^2)$ for $n$ noisy observations of the true function.

\smallskip

\section*{Appendix}


\subsection{Proof of Theorem~\ref{th.interpolation}}\label{app.prof}

The theorem follows from the discussion in \cite[Remark~2]{Taylor2016} with some extra computations. For the sake of completeness, we report it here. In particular, take any set $\{(\tilde{\x}_i,\tilde{\g}_i,\tilde{f}_i )\}_{i\in I_n}$. If this set is $\mathcal{F}_{1/(L-\mu), \infty}$-interpolable, then an allowed interpolating function is \cite[Remark~2]{Taylor2016}: 
$$
h(\tilde{\x}) = \max_{i}\{h_i(\tilde{\x})\}, 
$$
with 
\begin{equation}
h_i(\tilde{\x}) = \tilde{f}_i + \tilde{\g}_i^\transp(\tilde{\x}-\tilde{\x}_i) + \frac{1}{2(L-\mu)} \|\tilde{\x} - \tilde{\x}_i\|_2^2.
\end{equation}
As in~\cite[Remark~2]{Taylor2016} by convex conjugation and curvature addition, an $\mathcal{F}_{\mu, L}$ interpolating function of $\{({\x}_i,{\g}_i,{f}_i )\}_{i\in I_n}$ is:
\begin{equation}
\hat{\varphi}_n(\x) = \conv(h_i^{\star}(\x)) + \frac{\mu}{2} \|\x\|^2_2.
\end{equation}
Computing $h_i^{\star}(\x)$ can be done in the standard way, 
\begin{multline}\label{hst}
h^{\star}_i(\x) = \sup_{\tilde{\x}} \{\x^\transp\tilde{\x} - h_i(\tilde{\x})\} =  \\ \sup_{\tilde{\x}}\{ \x^\transp\tilde{\x} - (\tilde{f}_i + \tilde{\g}_i^\transp(\tilde{\x}-\tilde{\x}_i) + \frac{1}{2(L-\mu)} \|\tilde{\x} - \tilde{\x}_i\|_2^2)\}.
\end{multline}
For the inner optimization problem (sup), by first-order optimality conditions, 
\begin{equation}
-{\x} + \tilde{\g}_i + \frac{1}{(L-\mu)} (\tilde{\x} - \tilde{\x}_i) = 0 \iff \tilde{\x} = \tilde{\x}_i + (L-\mu) ({\x} - \tilde{\g}_i),
\end{equation}
which substituted back into~\eqref{hst} yields
\begin{equation}\label{hst1}
h^{\star}_i(\x) = \frac{L-\mu}{2} \| {\x} - \tilde{\g}_i\|_2^2 + \x^\transp \tilde{\x}_i - \tilde{f}_i.
\end{equation}
From \cite[Theorem~4(c)]{Taylor2016} follows that 
$$
\{(\tilde{\x}_i,\tilde{\g}_i,\tilde{f}_i )\}_{i\in I_n}=\{(\g_i-\mu \x_i, \x_i, \x_i^\transp\g_i - f_i- \mu/2\|\x_i\|_2^2)\}_{i\in I_n}.$$
Operating the substitutions in~\eqref{hst1} and calling $p_i(\x) := h^{\star}_i(\x)$, the thesis follows. \hfill $\blacksquare$









\bibliographystyle{IEEEtran}
\bibliography{../../PaperCollection00}

\end{document}